\documentclass{elsarticle}

\usepackage{hyperref}
 \usepackage{color}
\usepackage{cancel}
\biboptions{sort&compress}    

\journal{Physica A}

\usepackage{natbib}
\usepackage{numcompress}
\begin{document}
\begin{frontmatter}

\title{Wealth distribution of simple exchange models coupled with extremal dynamics}
\author[DF]{N. Bagatella-Flores}
\ead{nbagatella@uv.mx}
\author[DF]{M. Rodr\'iguez-Achach\corref{mycorrespondingauthor}}
\ead{manurodriguez@mda.uv.mx}
\cortext[mycorrespondingauthor]{Corresponding author}
\author[DIA]{H.F. Coronel-Brizio}
\ead{hcoronel@uv.mx}
\author[DIA,DT]{A.R. Hern\'andez-Montoya}
\ead{alhernandez@uv.mx}
\address[DF]{Universidad Veracruzana. Facultad de F\'isica e Inteligencia Artificial. Departamento de F\'isica, Lomas del Estadio s/n Zona Universitaria, Xalapa, Veracruz 91000. M\'exico}

\address[DIA]{Universidad Veracruzana. Facultad de F\'isica e Inteligencia Artificial. Departamento de Inteligencia Artificial. Sebasti\'an Camacho 5, Xalapa, Veracruz 91000.  M\'exico.}

\address[DT]{Programa de Doctorado Sobre Desarrollo Cient\'{\i}fico y Tecnológico para la Sociedad (DCTS).\\
Centro de Investigaci\'on y de Estudios Avanzados del IPN. \\
Av. Instituto Polit\'ecnico Nacional 2508, Col. San Pedro Zacatenco,\\
Delegaci\'on Gustavo A. Madero, C\'odigo Postal 07360
Apartado Postal: 14-740, 07000 M\'exico, D.F., M\'exico.}

%
%
%

\begin{abstract}
Punctuated Equilibrium (PE) states that after long periods of evolutionary quiescence, species evolution can take place in short time intervals, where sudden differentiation makes new species emerge and some species extinct. In this paper, we introduce and study the effect of punctuated equilibrium  on two different asset exchange models: The yard sale model (YS, winner gets a random fraction of a poorer player's  wealth) and the theft and fraud model (TF, winner gets a random fraction of the loser's wealth). The resulting wealth distribution is characterized using the Gini index. In order to do this, we consider PE as a perturbation with probability $\rho$ of being applied.  We compare the resulting values of the Gini index at different increasing values of $\rho$ in both models.
We found that in the case of the TF model, the Gini index  reduces as the perturbation $\rho$ increases, not showing dependence with the  agents number. 
While for YS we  observe a phase transition  which happens around $\rho_c=0.79$. For perturbations $\rho<\rho_c$ the Gini index  reaches the value of one as time increases (an extreme wealth condensation state), whereas for perturbations bigger or equal than  $\rho_c$ the Gini index becomes different to one, avoiding the system reaches this extreme state. We show that both simple exchange  models coupled with PE dynamics give more realistic results. In particular for YS, we observe a power low decay of wealth distribution.
\end{abstract}
\begin{keyword}
Econophysics \sep  Agents exchange models \sep  Punctuated equilibrium \sep Wealth distribution \sep Gini index 
\PACS 89.75.-k \sep 05.65.+ \sep 05.50.+q \sep S89.65.Gh 
\end{keyword}
\end{frontmatter}

\section{\label{sec:intro}Introduction}

In the seminal paper of Bak and Sneppen, a Self-Organized Critical (SOC)
model \cite{Bak} is introduced to describe the ecological co-evolution of
interacting species. The success of the model in reproducing the punctuated
equilibrium behavior proposed by Gould \cite{Gould}, and already observed 
in the fossil records \cite{Raup}, has attracted many authors to study the 
model, and variations of it, through several approaches ranging from 
simulation \cite{Grassberger},\cite{Rios}
to the renormalization group \cite{Marsili}, \cite{Mikeska}. Bak and Sneppen's
model has found interesting  applications in economic studies \cite{Ausloos}, \cite{Yamano}, bacterial evolution \cite{Bose} and even optimization
problems \cite{Stefan},\cite{Boettcher}.

The Theory of Punctuated Equilibrium emerged as an opposition to Phyletic gradualism, which  is a theory of speciation that states evolution occurs uniformly and by the steady and gradual transformation of whole lineages, so that no clear line of demarcation exists between an ancestral species and a descendant one.  Punctuated equilibrium, on the contrary, states that evolutionary change takes place in short periods of time tied to speciation and extinction events, separated by large time periods of evolutionary quiescence, called stasis.  Evidence for these ideas has been found in the fossil record of bryozoans \cite{Cheetham}. This record shows that the first individuals appeared about 140 million years ago, remain unchanged for its first 40 million years (stasis). After that,  an explosion of diversification is observed, followed by another period of stasis. Other well known events, observed in the fossil record and explained by Punctuated Equilibrium are the extinction of dinosaurs, about 50 million years ago, or the huge number and sudden emergence of new species during the Cambrian period, in the Paleozoic Era, about 500 million years ago, called the Cambrian Explosion.

On the other hand, the study of  wealth and income distributions in society, is a very important and fundamental area of research for practical and theoretical reasons to social scientists, economists, econophysicists, sociologist, philosophers, etc. and also concerns to politicians, government administrators, international bankers,and surely to  national security agencies from many countries and of course to every common citizen. Although questions on the origin and causes of inequality are very old; attempts to answer them have been not very successful, even if many ideas have been proposed to understand and solve the problem. Between these ideas, we can mention the following:  difference in religious ethics, lack of a  qualified workforce, dependence on external technology, low level of internal savings, non-equilibrium between exports and imports, low cognitive and schooling skills of population, level of corruption and quality of democracy, capital's rate of return exceeding rate of output and income,  and many more \cite{Weber, Ranis,Lewis,Chenery,Taylor,Hanushek,Hilman,Hongyi,Acemoglu, Piketty}.

Even more, large scale social and ideological experiments, intended and implemented by force, to solve the inequality problem by centralization of economy, have failed spectacularly with a terrible prior and posterior cost in human  suffering and lives, human rights violations, famines, waste of economic resources, political and economical instability, ``hot'' and ``cold'' wars, immigration waves, etc.

The important fact is that currently the extreme economic inequality problem seems is not any more only  restricted  to the beforehand called ``Third world countries'',  but is  also becoming a big concern and serious  problem in developed economies, where the social and wealth gap between the low-medium income segments of population  and the richer one, has been recently increasing fast and systematically~\footnote{Although some economists and policy designers do  not make any distinction between inequality and poverty, they  are different issues. A society or country can be quite equal with a high number of very poor people or vice versa.} (for an extensive and polemic discussion on this topic see \cite{Piketty}).   

 The first empirical studies to understand wealth distribution were made by Pareto \cite{pareto}, who proposed that  the wealth and income distributions  obey an universal power law. Subsequent studies have shown  that this is not the case for the whole range of population  wealth values. Mandelbrot \cite{mandelbrot} proposed that the Pareto conjecture  only holds  at the higher values of wealth and income. The initial part  (low wealth or income) of the distribution has been identified  with the Gibbs distribution \cite{chakrabarti3,chakrabarti5},  while the  middle part,  according to Gibrat \cite{gibrat}, takes the form of a log-normal  distribution.  

 Recently, due to great  advances in Complexity Sciences and computing power  new ways to model and understand social and economic systems have emerged. Between the most important and well known applications of this computational methods  we can mention the use of  multi-agent based models to investigate the problem of  wealth distribution \cite{chakrabarti5,Dragulescu,Bouchaud,Chatt,scalas,scalas3}

In this work, by using  a multi-agent computer methodology, we explore the effect of introducing the extremal dynamics of the already mentioned Bak-Sneppen model,  on the wealth distribution produced by two very simple economic exchange models and study their corresponding Gini indices.

 In particular, we focus our attention in two well known toy-models of economic interactions that have been used extensively due to their simplicity, such as the so called ``Yard-Sale" (YS) and
``Theft and Fraud" (TF) models \cite{Hayes}. Although these two models have the 
advantage of their simple rules for analysis and simulation, they are over simplified, toy model versions of a real economy and they do not produce realistic wealth distributions. For this reason, several authors have made some refinements to introduce and model more realistic situations, such as the introduction of savings \cite{chakrabarti}, changing the probability of winning according to the 
relative wealth of the traders \cite{Sihna}, allowing the agents to go into debt \cite{Dragulescu} and by the introduction of altruistic behavior \cite{Trigaux, achach}. Interesting and deep textbook discussions of multi-agents exchange models in the context of the present  work are \cite{Scalasbook} and \cite{Chakabook}.

\section{Model Description}

We can treat  an economy in its simplest form as an interchange of wealth 
between pairs of economic entities (people, companies, countries, etc.), named our ``agents" at successive instants of time. Every time  two agents interact, wealth flows from  one to the other  according  to some rule. In the case of the YS model, 
the winner takes a random fraction of wealth from  the
poorer player, while in the TF case, 
the winner takes a random fraction of the loser's wealth.
There is not production or consumption of wealth in these models, 
no taxes, savings, etc. Under these circumstances, the YS model produces a
collapse in the economy: all the wealth ends  in the hands of a single 
agent, a phenomenon know as extreme wealth condensation. On the other hand, the TF model 
does not collapse, but leads to a wealth distribution given by the Gibbs 
distribution~\cite{Dragulescu1,chakrabarti5,ispolatov}.  In particular,  the introduction of punctuated equilibrium in these two  
models has not been  studied in depth, and therefore, in this paper we 
investigate the effect that punctuated equilibrium  behavior  has on the 
dynamics  of the models  and the changes  that it can produce on the 
distribution of the wealth.

As discussed in \cite{Hayes}, after a sufficient number of interactions take place in the YS model, a single agent ends up with (almost) all the money  in the system.
On the other hand, the TF model produces a distribution with the 
majority of the agents ending with a wealth close to the average, and no 
agent becomes extremely rich.
Note that in the TF case, a very poor agent that interacts with a rich one, 
can become rich if he wins the bet and the fraction is enough high, a situation that is not expected to happen in the real world, unless we are dealing with illegal activities, hence the name of the model. 

These two outcomes from the YS and TF models actually do not reflect what 
really happens in a modern economy, where the wealth distribution takes 
an exponential  distribution form for the poor and medium class sectors
 of the population, and a Pareto distribution for the richer individuals 
\cite{Estevez,Wright}.

\subsection{Simulation implementation}
Our simulation runs in the following way: $N$ agents are  arranged on an 
one dimensional lattice with periodic boundary conditions. This lattice is not important in any trading activity between agents, however it will be very important and  necessary to keep track of every agent's first neighbors, in order to apply EP rules to introduce wealth mutations in a subsequent step of our simulation as explained below.  Initially a certain  amount of money $M$,  is distributed equally to all agents in such a way that 
$\Sigma_{i=1}^N m_i= M$. The system is closed, meaning that the total amount of money in the system, $M$, is always constant (i.e. no production) and the number of agents $N$ remains unchanged (neither dead, birth or  migration of agents is allowed). At each time step, two agents are randomly chosen and they interact according to the YS or the TF model. However, with probability $\rho$, the interaction will be ruled by punctuated equilibrium (PE) dynamics instead of YS or TF.  Then, another pair of agents are chosen and the process is repeated $K$ times,  which constitutes one Monte Carlo step (MCS). In our simulations and to obtain wealth distributions showed in next section a  typical number for $K = 10^6$  and every agent starts the simulation owning 100 monetary units.

Punctuated equilibrium (PE) is introduced in both  models in the following 
way: locate  the poorest agent, lets say agent $k$ and assign new values of wealth to agents $k-1$, $k$ and $k+1$ at random, but taking care that their combined wealth does not change. Extinction of agents  is not allowed  in order to maintain the overall number of agents $N$ constant.

In the case of the YS or TF rules, two agents $i$ and $j$ are randomly 
chosen at time $t$. The winner, which is also chosen at random, takes an amount $T$ 
of money from the loser. 
The traders' wealth $w_i$ and $w_j$ at time $t+1$, assuming that agent $i$ is the winner and agent $j$ is the loser, will be
\begin{eqnarray}
& & w_i(t+1)=w_i(t)+T\ ,\\
& & w_j(t+1)=w_j(t)-T\ ,
\end{eqnarray}
Where $w_j$ is the wealth of the poorest agent for the YS model and $w_j$ is the wealth of the loser for the TF model.

\noindent
The amount $T$ of the wealth that changes hands in the bet is defined as

\begin{equation}
T = \alpha MIN (w_i(t), w_j(t)),
\end{equation}

\noindent
for the YS model and

\begin{equation}
T = \alpha w_j(t),
\end{equation}

\noindent
for the TF model, assuming that agent $j$ is the loser, where the parameter 
$\alpha$ is a random number from an uniform distribution 
in the interval [0,1]. 

The inequality in the final wealth distribution of the system can be quantified using the Gini index
\cite{Berrebi}, defined by

\begin{equation}
G=\frac{\Sigma^N_{i=1} \Sigma^N_{j=1} |x_i-x_j|}{2N^2 \mu},
\end{equation}

\noindent
where $\mu$ is the average wealth, and $x_i$ and $x_j$ represent the wealth 
of agents $i$ and $j$ respectively. A perfect distribution  of wealth, where 
everybody has the same amount of money will give a value $G=0$. The other extreme, where one individual owns all the money has a Gini value of 1. 

\section{Simulations Results}
\subsection{Gini Index Analysis}
\begin{figure}[htb!]
\begin{center}
\resizebox{0.80\textwidth}{!}{%
\includegraphics[angle=-90]{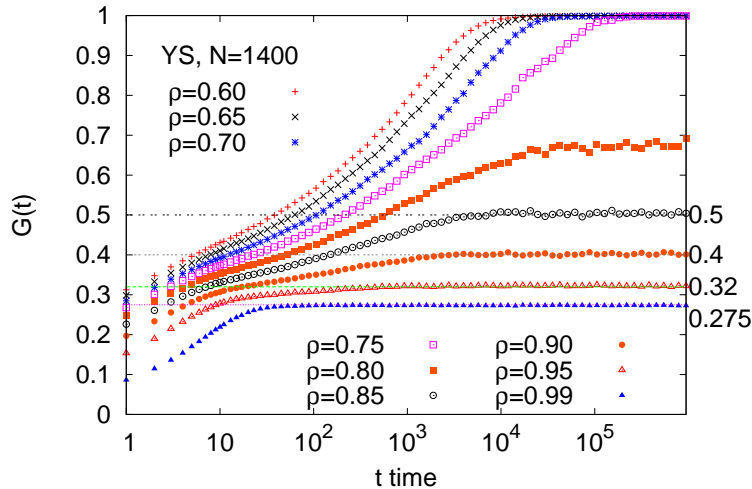}}
\caption{Gini index $G(t)$ as a function of time for the YS Model for
$N=1400$ agents. We set up the strength  of the perturbation $\rho$ to the values $\rho$=0.6, 0.65, 0.7, 0.75, 0.8, 0.85, 0.9, 0.95, 0.99. In this 
figure the upper curve corresponds to $\rho$=0.6, and the bottom one 
corresponds to $\rho$=0.99. We observe that as $\rho$ increases, the curves tend to reach the $G =1$ region  more slowly. For  $\rho \ge  0.8$ the curves avoid completely that region.}
\label{giniYS1400}
\end{center}
\end{figure}

\begin{figure}[htb!]
\begin{center}
\resizebox{0.80\textwidth}{!}{%
\includegraphics[angle=-90]{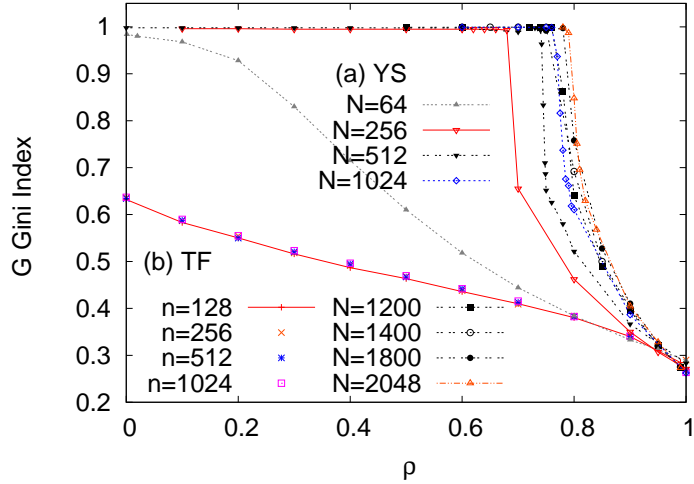}}
\caption{Gini index  $G(p)$ as a function of perturbation for our two models: (a) YS model: $G(\rho)$ curves plotted for a different increasing number of agents $N$ indicated in the figure. (b) TF model: In this case, we use lower case $n$ to indicate the  number of agents. $G(\rho)$ curves were plotted for $n$=128, 256, 512, 1024 values.}
\label{bothGiniI}
\end{center}
\end{figure}

We first consider the Gini index in the YS model as a function of time, for several
values of the PE ``perturbation" $\rho$. The results are shown in Figure \ref{giniYS1400}, where one can see that for  small enough values of $\rho$, the final result is the same as with the ``pure" YS model, that is, $G(t)$ reaches the value of 1 as time increases (the economy collapses in a state where a single agent has all the money). However, there is a critical value of $\rho_c$ for which the system does not collapse and $G$ takes an asymptotic value less than 1. As the system size increases, the perturbation necessary to get the system out of collapse increases, as can be seen in  Figure \ref{bothGiniI}, where the asymptotic value of $G$ is shown as function of $\rho$ for several system sizes $N$. Same figure shows the results for the TF model. In this model the effect of the perturbation only decreases the asymptotic value of $G$.

\subsection{Wealth Distribution}
\begin{figure}[htb!]
\begin{center}
\resizebox{0.80\textwidth}{!}{%
\includegraphics[angle=-90]{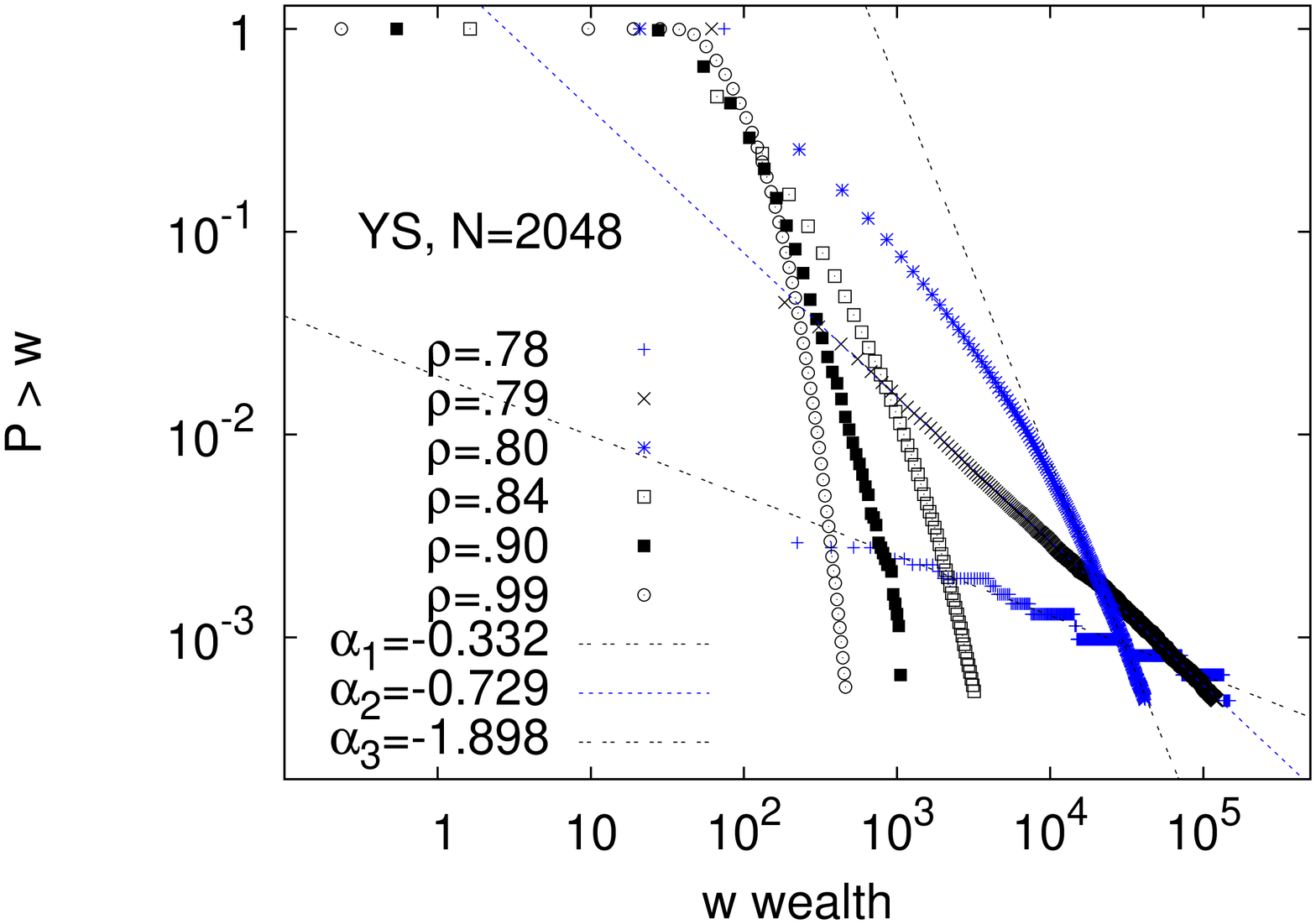}}
\caption{Wealth Cumulative Distribution Function (CDF) in a log-log  scale for the YS model. i.e. the probability $P$ that an agent chosen at random will have a wealth greater than or equal to $w$. We show simulations with $N=2048$ agents and different increasing values of $\rho \geq \rho_c$. We observe that as $\rho$ increases, wealth distribution becomes fairer. Power law fits for curves  determined by $\rho = 0.78$, 0.79 and 0.80  are displayed. Corresponding exponents are $\alpha_1$ = 0.332, $\alpha_2$ = 0.729 and $\alpha_3$ = 0.80. From second fit, we select  $\rho_c$ = 0.79. See table \ref{tab:PowerFits}. Curve for $\rho_3$ = 0.80 decays asymptotically as a power law fit, and it is the closer to the observed in real data.} 
\label{cumulative2048}
\end{center}
\end{figure}

Figure \ref{cumulative2048} shows in a log scale  for the YS model
the  wealth cumulative distribution function (CDF), that gives
the probability $P$ that an agent chosen at random will have a wealth greater than $w$. Here, $N=2048$ agents and were simulated the values for the perturbation $\rho$=0.78, 0.79, 0.80, 0.84, 0.90, 0.99.

From this figure \ref{cumulative2048}, we can see that, as the perturbation $\rho$ increases, the probability of finding richer agents decreases, giving us  a ``fairer'' distribution of wealth, compared to lower perturbations where only higher  values of wealth are found. 

We also observe in same figure that for the values of $\rho=0.78$ and $0.79$ it is possible to fit very well these curves with a power law model with the form $F(w) = cw^{-\alpha}$. Parameters $c$ and $\alpha$  of these  fits are displayed in table \ref{tab:PowerFits}

\begin{table}[!htb]
\begin{center}
\caption{Wealth CDF power law  fit parameters for values of $\rho= 0.78$ and 0.79. NDF denotes the number of degrees of freedom.}
\label{tab:PowerFits}  
\begin{tabular}{lccccr}
\hline
\hline
$\rho$ &$c$  &$\alpha$ & $\chi^2$/NDF& Selected Fit \\
\hline
\noalign{\smallskip}\hline\noalign{\smallskip}
0.78 &0.029 $\pm$ 0.062  &0.332  $\pm$ 0.062 &32.49/998& No\\
0.79 &  2.550 $\pm$ 0.024& 0.729  $\pm$ 0.002 &  4.956/998& Yes\\
\hline
\hline
\end{tabular}
\end{center}
\end{table}

Again, from figure \ref{cumulative2048} and value of $\chi^2$/NDF in table \ref{tab:PowerFits}, we can see that the best power law fit corresponds to $\rho_c$ = 0.79, with a  exponent $\alpha$ = 0.729. This confirms our observation of a phase transition around $\rho_c = 0.79$ described by the analysis of the behavior of the Gini index as a function of time for different  $\rho$ values and displayed in figure \ref{giniYS1400}.

For $\rho = 0.80$ the corresponding wealth CDF decays asymptotically as a power law with an exponent $\alpha_3$ = 1.898 $\pm$ 0.002. The fit was performed in the region $w > 14764$ on the biggest 647 observations $\chi^2$/NDF obtained is 0.1685/645. This exponent has a value enough close to the observed in real wealth cumulative distribution data that is approximately 2~\cite{Dragulescu1}
. Interestingly, the more realistic wealth distribution obtained does not corresponds to the value of $\rho_c$ (a pure power law distribution), but to a value close to it, $\rho=0.80$ which is an  like real asymptotic  power law distribution with an Pareto exponent close to, but slightly lower than 2.

\subsection{System Size effect}
\begin{figure}[htb!]
\begin{center}
\resizebox{0.80\textwidth}{!}{%
\includegraphics{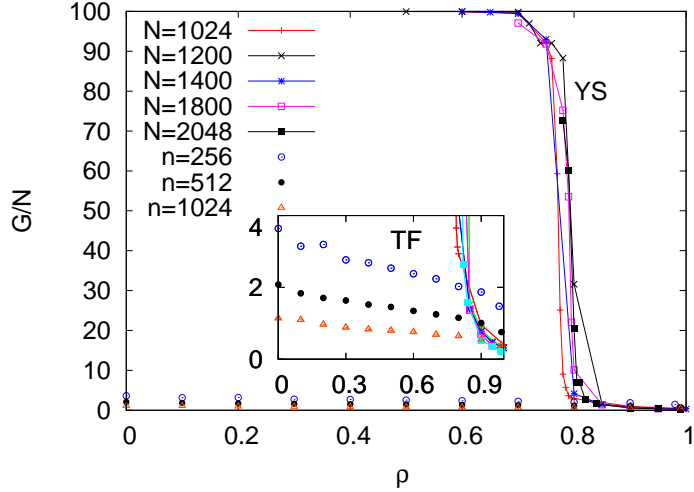}}
\caption{Gini index divided by the agents number N and scaled to 100 as a function of the perturbation  for  our two models:  (a) YS model: figure shows this scaled Gini index for five different number of agents $N$ with values indicated in the figure (b) TF model: Scaled Gini index for three different number of agents$ n$=256, 512, 1024 in the inset. Again we have used lower case $n$ to indicate the number of agents for TF case.}
\label{linescumulaN}
\end{center}
\end{figure}

After being established that the maximum wealth depends on the perturbation, we now proceed to investigate the effect of the system size on our simulations. This is shown in Figure \ref{linescumulaN}. We divided the Gini index from Figure 2 by the system size $N$ and then normalized it to 100. We can see that its behavior is independent of $N$.

\begin{figure}[htb!]
\begin{center}
\resizebox{0.85\textwidth}{!}{%
\includegraphics{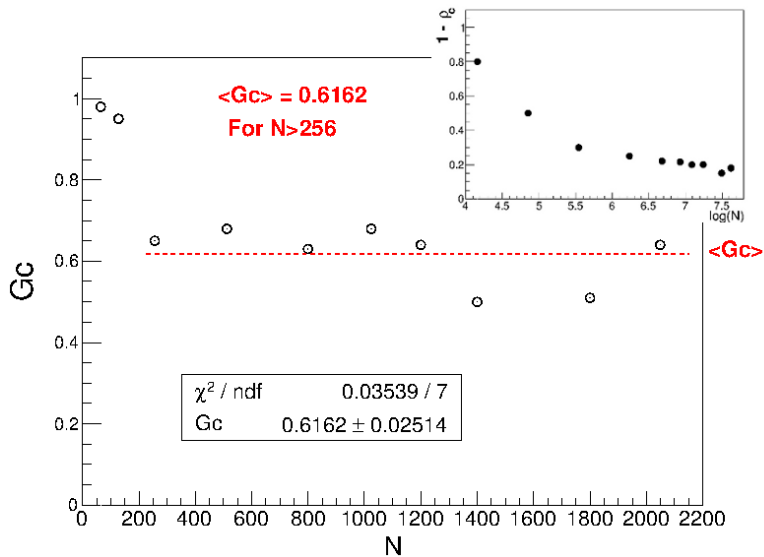}}
\caption{YS model. Gini index  threshold  $G_c$ as a function of the number of agents $N$. We can see that $G_c$ seems to become independent of $N$.  For $N > 256$ It fluctuates around its mean value $<G_c>$. Red broken line was obtained by a $\chi^2$ fit procedure applied to the 8 last data points. We also  show in the inset the plot $(1-\rho_c)$ vs $log(N)$. It can be observed that $(1-\rho_c) \rightarrow 1$ very slowly and asymptotically}
\label{cumulativeC2048}
\end{center}
\end{figure}

For the YS model, in Figure  \ref{cumulativeC2048} we show that the  critical Gini index $G_c$, does not depend on the number of agents $N$ involved in our simulations. $G_c$ fluctuates around a mean value of  $<G_c> = 0.6162 \pm 0251$. The mean value $<G_c>$ was obtained by fitting a constant horizontal straight line to our data of $G_c$ values for a different number $N$ of agents. Same numerical value results from averaging directly our $G_c$ values until the fourth position after the decimal point ($10^{-4}$). This is a good result because if $G_c$ were dependent on $N$, the  wealth distribution for the YS model should change on the number of agents, something that should not happen in a good simulation

Continuing with the YS model, in the upper inset of same Figure \ref{cumulativeC2048} we plot the critical  $\rho_c$ as a function of $log(N)$. We can see as was pointed out above, at the beginning of this section, that the critical perturbation increases very slowly  with the system size. In fact, since we have defined $\rho$ as the probability of switching on or not PE in the  interactions between agents, its numerical value can not be larger than one.  For the extreme case of``an infinite system size'' we should have  $\rho_c\rightarrow 1$, case where  nothing can be done to avoid that the system collapses increasing the strength of the perturbation $\rho$. Of course this extreme case does  not represent any  real economic system, since that although they  can be constituted of a very big number of agents, their number is always finite.

\section{Conclusions}

In this work, we analyze the TF and YS models under the influence of the  punctuated equilibrium dynamics, which is introduced  as a perturbation $\rho$ determined by the probability of applying or not PE in agents exchange of money. Although TF model is weakly affected by the introduction of PE, for the case of YS and for perturbations $\rho >0.8$ , the asymptotic Gini index becomes different to one, meaning that the perturbation avoids the collapse in the economy where a single agent takes all money, re-allocating  it between agents  in a way that a lower inequality in the distribution of wealth is observed. Even more, a phase transition around a critical value $\rho_c = 0.79$ is observed . For this critical value of $\rho$, the corresponding wealth distribution displays a power law decay with an exponent $\alpha=0.729 \pm 0.002$. For a value of $\rho$ = 0.80 the wealth distribution decays asymptotically as a power law with an exponent $\alpha_3$ =  1.898 $\pm$ 0.002, consistent with the values observed in real data.

The resulting wealth distributions can be tuned to different values of the Gini index, adjusting the perturbation. A corollary from this, that is possible to extend to real economics  is that a fairer wealth distribution is only attainable by means of a mechanism of wealth re-distribution, as by example taxes in real life. The implementation of different re-distribution mechanisms or determining what  is the optimum proportion of individual income to taxing or re-distribute are problems that also could be explored through agents modeling methodology.

Even if the evolutionary economic  approach presented here seem very naive and far from representing real economic phenomena, it  has precisely the virtue that  preserving the simplicity of YS and TF models, produces results more realistic than the obtained by the original models without  PE dynamics included, such as a finite wealth distribution decaying as a power law. This is important for the  agents model theory and implementation goal of constructing a minimum agents model for real economic and social systems. Also, and in our opinion, our results are important for the most ambitious  and long term end of constructing a real microeconomics theory, in the sense of statistical physics, if possible \cite{Scalas2}. We believe that an initial and main way of attacking successfully this difficult problem is through the intensive use of multi-agent simulations techniques, followed by a formalization of results and techniques emerged from  this approach. For a nice discussion on these issues in the context of diverse agents models and SOC, see \cite{Jeldtoft}, chapter 5.

Besides, other studies have been made where the simple YS and TF models are extended including effects of savings, of course taxes and other mechanisms in order to make them more realistic. We believe that our approach using the extremal dynamics of the Bak-Sneppen model which has the effect of re-allocating wealth between agents, is simpler and yields similar results, a feature that can be used to investigate particular economic phenomena with a simpler model. This can benefit both simulation and analytic studies. Any way, the study of many-body real world systems, through the study of computational models is not an easy task, considering that sometimes we do not understand in deep those models. In our case, it would be very interesting to attack the same problem presented here using a more formal approach as a Mean Field Model. This is matter of a future work.

\section*{Acknowledgments}
This work has been supported by SEP-Conacyt (M\'exico) under project grant numbers 155492 and 135297. We also thank PROMEP and SNI (M\'exico) for continued support. Some of the analyses of this paper were performed using ROOT \cite{root}. We thank pertinent comments and suggestions of E. Scalas.


\bibliography{eqpuntuado2014FV}

\end{document}